\documentstyle[11pt,paspconf]{article}

\begin{document}     
\title{ Variability Models of Gamma-Ray Blazars}

\author{Amir Levinson}
\affil{School of Physics and Astronomy, Tel Aviv University, 
Tel Aviv 69978, Israel}

\section{General remarks}
Much observational efforts have been devoted in recent years to 
study blazar variability across the electromagnetic spectrum 
(for a recent review see e.g., Ulrich, et al. 1997).  In 
addition to spectral and 
polarization information, variability data can provide stringent 
constraints on the radiation mechanisms, the geometry of the 
emission regions, beaming factors, formation and dynamics of 
shocks, and perhaps the jet's content. 

In most models the variability pattern is governed by the 
following timescales: i) the light travel time across the 
source, ii) the cooling time, iii) the acceleration or 
injection time of radiating particles, and iv) the
dynamical time, which equals roughly the light crossing time in the
case of a relativistically expanding source. 
It is conceivable, however, that the temporal structure observed 
involves  additional, distinct timescales that are associated
with completely different physical process, as demonstrated by other 
transient systems, e.g., radio pulsars, GRB; the overall pulse duration
and the duty cycle in the former system reflect the 
rotation of a neutron star, whereas the temporal substructure 
(sub-pulses, polarization swings, etc.) is presumably connected 
with the emission mechanism.  A plausible 
variability scenario for blazars is the formation of a 
train of shocks during a period of enhanced activity that 
might be associated with accretion instabilities or 
with the process responsible for the ejection of the jet.   
Such a possibility 
seems to be suggested by some recent observations which reveal, 
what appears to be rapid flaring during the occurrence of a much 
longer outburst (Wagner 1998).

In models whereby the emission originates from deep inside the jet,
the size of the source is limited by optical depth effects.
The gamma-spheres in the powerful sources and the 
photospheres of the IR-to X-ray emission lie in the range between
$10^{-3}$ to about 1 pc (Blandford \& Levinson 1995, and Levinson 
1996).  The corresponding light travel times, as measured
by a distant observer at small viewing angles, then range 
from a few minutes to several weeks, assuming that the emitting
plasma moves with Lorentz factor $\Gamma\simeq 10$.  The 
radio-spheres are typically located at much larger radii.
This range of time scales is in accord with the rapid variability
frequently observed in blazars (IDV has been observed
in many bands, with changes on time scales as short as a few minutes
in the optical and X-ray bands and a few hours in gamma-rays;
Wagner 1997 and references therein).  It also illustrates
the temporal resolution and sampling rates required for testing 
predicted correlations or other model features.

Correlations between optical and gamma-ray emission in flat spectrum
radio quasars, and between X-ray and TeV emission in BL Lac objects 
appears to be another characteristic of blazar variability.  The 
time lags between different
bands and the relative amplitudes are important diagnostics of 
the radiation mechanisms and the structure of the emission region.
They may also be influenced by geometrical and 
orientation effects (see below).  Unfortunately, it seems that 
the sensitivity available, particularly in the gamma-ray
band, is insufficient to provide good enough time resolution to 
test relevant model predictions.  
Another caveat is that the time separation between subsequent 
observations in recent multiwaveband campaigns (e.g., the 
high and low states in the 1994 campaign on  3C279; Maraschi 
et al. 1994, or the pre-flare and high state in the 1996 
campaign ; Wehrle et al. 1998) is very long compared with 
the variability time anticipated.  Therefor, conclusions 
regarding the radiation mechanism for instance, which are drawn based 
on data taken at say two epochs in some individual source can be 
misleading.  A better strategy might be to look for 
systematic trends (e.g., delays between gamma-ray and radio
outbursts, as predicted by inhomogeneous models, or changes of cutoff
energies) in a sample of sources. 
It is hoped that the next generation
gamma-ray telescope and forthcoming campaigns will help 
elucidating the relation between the emission in different 
bands, and discriminating between models.

\section{Models of blazar variability}

Several types of variability models have been discussed in the 
literature.  In one class of models, some of the source parameters (e.g.,
magnetic field, density) and/or particle acceleration rate
are assumed to have explicitly time dependence.  Such models may 
represent a physical situation in which   
sudden changes of the outflow parameters and/or 
particle acceleration rate result from e.g., magnetic 
reconnection episodes, as in the case of solar flares, encounter
of a strong shock with a region of enhanced density or magnetic field,
or various types of instabilities.  The time dependent SSC model 
by Mastichiadis \& Kirk (1997) and Kirk, Rieger \& 
Mastichiadis (1998), who applied it to 
the TeV BL Lac objects, is an example.
In ERC models the variability can be produced also by changes 
of the ambient radiation intensity.    
Geometrical effects may also have important implications for the
observed variability.  In particular, rapid variations of the
observed flux in any band can be produced without straining the 
parameters of the emission region too far (Salvati, et al.  1998).  
Such effects may provide an explanation for the radio IDV
which is problematic for other models (Wagner 1998).  
Another class of models associates the temporal behavior 
of blazars with the dynamics of shocks or blobs (e.g., Dermer \& 
Chiang 1998; Levinson 1998a).  Here the 
variability is produced by implicit time changes of the blob
or front parameters that are associated with the inhomogeneity of 
the source.  In the following we discuss a particular model of 
this type in some greater detail.

In the radiative front model (Romanova \& Lovelace, 1997;
Levinson 1998a) the variable emission
seen originates inside dissipative fronts that are
produced by overtaking collisions of highly 
magnetized, relativistic outflows, and consist of a pair of 
shocks and a contact discontinuity.  In the regime where the ERC 
process dominates the production of the high-energy 
emission, the shape and timescale of the flare depend on the ratio
of the thickness of expelled fluid slab and the gradient length
scale of background radiation intensity; when this ratio exceeds unity
then the shape of the light curve is determined by 
the radial variation of ambient radiation intensity,
and is typically asymmetric with a rapid rise
and a longer decay.  When it is much smaller than unity, the 
flare duration is determined by the shock travel times across the 
fluid slabs (the cooling time is typically shorter).  In this 
case the decay is comparable to or shorter than the rise.
Since the ERC emission is anisotropic in the front frame it 
gives rise to a radiative drag and consequent
deceleration of the front during the rise of the radiated flux.  This
renders the amplitude of variations and the high-energy cutoff of the 
emitted spectrum sensitive to the Thomson opacity.  Depending upon
the conditions in the source, pair production effects can lead 
to either, a high-energy cutoff in the emitted spectrum or
a propagating flare with longer delays, longer durations and 
smaller amplitudes
for higher energy gamma-rays.  In view of the relatively short
timescales involved (see \S 1) and the decrease in amplitude 
with increasing gamma-ray energy, the detection of such delays in the
gamma-ray band requires good sensitivity, typically much better than 
provided by EGRET, and should be one of the prospects of 
future gamma-ray missions. 
Since the synchrotron flux at low frequencies is self-absorbed at 
radius of peak emission, radio outbursts lag gamma-ray flares 
in this model.  Such delays have been observed in several cases
(e.g., Wehrle et al. 1998; Otterbein et al. 1998).  Detailed account
of the correlations predicted for the angle averaged flux 
will be given elsewhere (Levinson, in prep.).  The radiative feedback should 
also give rise to a dependence of the variability on the orientation 
of the source (Levinson, 1998b).  Apart from changing the shape of
the light curves, it can significantly affect the predicted 
correlations.  For example, the radio flux is emitted after 
the front re-accelerates to its initial velocity and is, therefore,
more beamed than the flux emitted during the peak (at higher 
energies).  Consequently, if viewed at
angles larger than the beaming cone of the radio
emission but smaller than that of the gamma-ray emission, 
a source may exhibit events whereby high-energy outbursts are 
followed by a small or no change of the radio flux.  There is 
some observational evidence
for such events (Mattox, priv. communication).

\section{Acknowledgment}

Support by Alon Fellowship is acknowledged.

\end{document}